\documentclass{pasj00}
\begin{document}
\SetRunningHead{NAKANISHI et al.}{Outer Rotation Curve of the Galaxy with VERA II: Annual Parallax and Proper Motion of IRAS21379+5106}
%\Received{}%{yyyy/mm/dd}
%\Accepted{}%{yyyy/mm/dd}
%\Published{}%{yyyy/mm/dd}

\title{Outer Rotation Curve of the Galaxy with VERA II: Annual Parallax and Proper Motion of the Star-Forming Region IRAS21379+5106}
\author{%
   Hiroyuki \textsc{Nakanishi}\altaffilmark{1}
   Nobuyuki \textsc{Sakai}\altaffilmark{2}
   Tomoharu \textsc{Kurayama}\altaffilmark{3}
   Mitsuhiro \textsc{Matsuo}\altaffilmark{1}
   Hiroshi \textsc{Imai}\altaffilmark{1}
   Ross A. \textsc{Burns}\altaffilmark{1}
   Takeaki \textsc{Ozawa}\altaffilmark{1}
   Mareki \textsc{Honma}\altaffilmark{2,4}
   Katsunori \textsc{Shibata}\altaffilmark{2,4}
   and 
   Noriyuki \textsc{Kawaguchi}\altaffilmark{2,4,5}
}
 \altaffiltext{1}{Graduate Schools of Science and Engineering, Kagoshima
   university, 1-21-35 Korimoto Kagoshima}
 \email{hnakanis@sci.kagoshima-u.ac.jp}
 \altaffiltext{2}{Mizusawa VLBI Observatory, National Astronomical
   Observatory of Japan, Mizusawa, Ohshu, 023-0861}
 \altaffiltext{3}{Center for Fundamental Education, Teikyo University of Science, 2525 Yatsusawa, Uenohara, Yamanashi 409-0193}
 \altaffiltext{4}{Department of Astronomical Science, SOKENDAI (The Graduate University for Advanced Studies), Mitaka, 181-8588}
 \altaffiltext{5}{Shanghai Astronomical Observatory, Shanghai, China} 

%% `\KeyWords{}' always has to be placed before `\maketitle'.
\KeyWords{astrometry, H {\sc ii} regions, Galaxy: kinematics and dynamics, ISM: individual (IRAS 21379+5106), stars: formation} %Do NOT move this preamble from here!

\maketitle

\begin{abstract}
We conducted astrometric VLBI observations of water-vapor maser emission in the massive star forming region IRAS 21379+5106 to measure the annual parallax and proper motion, using VERA. 
The annual parallax was measured to be $0.262 \pm 0.031$ mas {corresponding to a trigonometric distance of} {$3.82^{+0.51}_{-0.41}$} kpc. The proper motion was $(\mu_\alpha\cos{\delta}, \mu_\delta)=(-2.74 \pm 0.08, -2.87 \pm 0.18)$ mas yr$^{-1}$. 
{Using this result}, the {Galactic} rotational velocity was estimated to be $V_\theta=218\pm 19$ km s$^{-1}$ at the Galactocentric distance $R=9.22\pm0.43$ kpc, when we adopted the Galactic constants $R_0=8.05\pm 0.45$ kpc and $V_0=238\pm 14$ km s$^{-1}$. 
With newly determined distance, {the bolometric luminosity of the central young stellar object was re-evaluated to $(2.15\pm 0.54)\times 10^3 L_\odot$, which corresponds to spectral type of} B2--B3. {Maser features were found to be distributed along} a straight line from south-west to north-east. In addition, a vector map of the internal motions constructed from the residual proper motions implies that maser features trace a bipolar flow and that it cannot be explained by simple ballistic motion. 
\end{abstract}

\section{Introduction}
VERA (VLBI Exploration of Radio Astrometry) is a Japanese VLBI {system} dedicated to exploring the three-dimensional Galactic structure, being operated by the National Astronomical Observatory of Japan and Kagoshima university \citep{hon00}. 
It consists of 4 stations, each with a 20-m antenna, located at Mizusawa,  Iriki, Ogasawara, and Ishigaki. 
In order to achieve high-precision astrometry, a dual-beam observation system is employed to observe Galactic maser sources and a distant QSO as a reference source, simultaneously \citep{kaw00}. Consequently, real-time phase calibration increases the signal-to-noise ratio and improves positional accuracy. 

The ``Outer Rotation Curve'' (ORC) project is one of the main projects carried out with VERA, which aims to {derive} the rotation velocity of the outer part of the Galaxy ($R > R_0$; $R$ denotes Galactocentric distance and subscript $0$ denotes the Sun). 
Since the inner rotation curve can be measured using {velocities at} the tangent points, it has been well determined \citep{bur78,fic89,mcc07}. However, in the case of the outer rotation curve, there are {no such useful points for} isolating the terminal velocity. Therefore, the distance of each outer Galactic source has to be measured directly. {As such}, in comparison with the inner Galaxy, there is much larger uncertainty seen {when compiling the} available outer rotation curves \citep{deh98a,sof09}.  
More efforts such as VLBI astrometry, are necessary to obtain a more accurate outer rotation curve {for the Milky Way Galaxy}. 

IRAS 21379+5106 is one of the ORC {project} targets, which is known to be an ultra-compact H {\sc ii} region \citep{bro96,wan09}. Since ultra compact H {\sc ii} regions are thought to be formed around massive stars, with a main sequence luminosity of B3 or hotter \citep{chu02}, IRAS 21379+5106 is considered to be a high-mass star-forming region \citep{wu06,she96}. 
A far-infrared, three color {(25, 60 and 100 $\mu m$)} image of the area around the source is shown in figure 1
%\ref{fig:RGBmap}
, where IRAS 21379+5106 is a point source indicated with a green circle. The position is $(\alpha, \delta)=(\timeform{21h39m40s.8}, \timeform{+51D20'35''})$ (J2000.0) in equatorial coordinates \citep{sun07}, which corresponds to $(l, b) = (95\fdg29670, -0\fdg93667)$ in Galactic coordinates. The radial velocity with respect to the Local Standard of Rest (LSR) is $V_{\rm LSR} = -42.27\pm 0.22$ km s$^{-1}$ from $^{13}$CO $ J=(2-1)$ observations \citep{wan09}, which we use as the LSR velocity of the source throughout this paper. IRAS 21379+5106 is thought to be associated with the Perseus arm, considering its position in Galactic coordinates and the LSR velocity \citep{dam01}, and the heliocentric distance has been estimated to be 5.9 kpc based on kinematice distance \citep{wan09}.

This source has been detected in some molecular lines such as CS($J=$2--1) \citep{bro96}, $^{12}$CO($J=$1--0) \citep{wou89,she96}, $^{13}$CO($J=$2--1) \citep{wan09}, H$_2$O maser \citep{wou93,bra94}, and NH$_3$ \citep{sun07}. \citet{wou89} noted that the $^{12}$CO($J=$1--0) spectrum showed a red shoulder profile, and the $^{13}$CO($J=$3--2) spectrum showed two peaks, with brighter emission in the blue-shifted component \citep{wan09}. {These line profiles can be interepreted as bipolar molecular outflows asscociated with IRAS 21379+5106.}

As Part II of the ORC project series, we report measurement of the annual parallax and proper motion of IRAS 21379+5106, observed with VERA, as well as the distribution and kinematics of the maser features. Section 2 describes details of the astrometric observations with VERA. Details of data reduction are summarized in Section 3. Section 4 presents our results, including: the annual parallax, trigonometric distance, proper motion, and maser distribution map. Section 5 discusses the 3-dimensional (3-D) motion of IRAS 21379+5106 in the Galactic disk, modification to the spectral type of the central young stellar object (YSO), and a bipolar outflow seen in the maser distribution. Section 6 reports the summary.

\section{Observation}
VLBI monitoring observations of IRAS 21379+5106 were carried out in 9 epochs from 2009 November 6 to 2011 November 5 in K-band with typical observation time of 8 hours. The observation frequency was set to 22.232--22.248 GHz to detect $6_{16} \to 5_{23}$ transition of the H$_2$O maser, which has a rest frequency of 22.23508 GHz. 

The observations were conducted in dual-beam mode \citep{kaw00}; one of the two beams, referred to as the A-beam, was pointed to the main target, IRAS 21379+5106, and the other beam, the B-beam, was pointed to the phase reference source J2137+5101. The tracking center was set to $(\alpha, \delta)=(\timeform{21h39m40s.55}, \timeform{+51D20'34''.0})$ (J2000.0). %, which was shifted from the formerly known position to point the better. 
The separation angle of the sources was $0\fdg52$ and both targets were observed simultaneously for phase calibration. 
The system noise temperature was typically 100--200 K throughout the observations. In addition to these two sources, we observed the bright quasar 3C454.3 with both the beams every 80 minutes for fringe finding and band-pass calibration. 

Other maser sources, IRAS 23385+6053 and IRAS 22555+6213, and continuum sources, J2339+6010 and J2302+6405, were also observed during these observation runs for other studies. 
During each observation run, each source was observed for an integration time of 10 minutes every 20 minutes. 
Hereafter we focus only on IRAS 21379+5106 since the other targets are outside the scope of this paper. 

The received signal was output into 16 {Intermediate frequency (IF) channels} after being filtered with the digital filter unit \citep{igu05}. The bandwidth of each IF was 16 MHz and {signals were} recorded with 2 bit sampling at a rate of 32 MHz. The total recording rate was 1024 Mbps. 
The correlation was performed with the Mitaka FX correlator, which outputted the auto-correlated and cross-correlated spectra with an interval separation of 1 second.

One IF was assigned to the A-beam, observing the maser line emission. 
The time-domain data was Fourier-transformed into spectra with 512 channels and a channel separation of 31.25 kHz, which corresponds to a velocity spacing of 0.4213 km s$^{-1}$. 
The other 15 IFs, each with a bandwidth of {16 MHz split over 64 channels, were} assigned to the B-beam, observing continuum emission from the reference sources. {During data reduction 14 of the continuum IFs were used and the 15th IF was abandoned due to a difference in correlator settings.} 

Observational settings and scheduling were compiled in the VEX (VLBI Experiment) format. 
Basic information for all the observations such as epoch, VEX file, Day-of-year (DOY) from 2009 January 1, synthesized beam, and noise are summarized in Table \ref{obssummary}.

\section{Data Reduction}
Data reduction was done with the software AIPS (Astronomical Image Processing System), developed by NRAO (National Radio Astronomy Observatory), which is commonly used in the data reduction of radio astronomy. 

The basic procedure of the primary reduction was as follows; (1) loading fits files (AIPS task: FITLD), (2) sampling bias correction using auto-correlation data (ACCOR), (3) a prior amplitude calibration (APCAL), (4) correction of delay-tracking, {which is necessary to replace the tracking model used in Mitaka FX correlator with more accurate tracking model calculated with CALC3/MSOLV \citep{jik05, man91}}, using a calibration table provided by the observatory (TBIN), (5) fringe finding of B-beam data (FRING), (6) applying the result of (5) to A-beam data (TACOP), (7) self calibration using B-beam data (IMAGR, CALIB, CLCAL), (8) applying the result of (7) to A-beam data (TACOP), (9) applying the calibration table {to correct the difference in path lengths of two receivers
in a dual-beam antenna with the "horn-on-dish method" \citep{hon08}} (TBIN), and finally (10) setting the velocity axis (CVEL). Further details of the primary reduction are explained in Appendix 1 of \citet{kur11}. 

After the primary reduction, we found three strong emission peaks detected in the cross-power spectrum, as shown in figure 2. 
%\ref{CPSPEC}. 
These three peaks were found at the LSR velocity ranges of $V_{\rm LSR}=-32.31$ {to} $-32.89$ km s$^{-1}$, $-42.00$ km s$^{-1}$, and $-50.43$ {to} $-46.63$ km s$^{-1}$ in every observation.   

Since the typical synthesized beam-size of VERA is 1 milli-arcsecond (mas) which is small compared to {the known position accuracy of $15''$ \citep{sun07}}, 
%the 139.6 - 152.6 arcsecond primary beam of each antenna (VERA status report 2011), 
we {needed to make} a fringe rate map \citep{wal81}, which enabled us to find the positions of maser spots with coarse resolution of $\sim 0.1''$ in order to reveal the general distribution of maser spots within the field of view. {Note that a `maser spot' refers to a point like emission region detected in a particular velocity channel.} 
Figure \ref{FRM} shows the fringe rate map made using the channel of $V_{\rm LSR}=-32.31$ km s$^{-1}$ {during} the first epoch.  
It shows that the position of the maser spots {are} shifted by $\Delta \alpha \cos{\delta}=-0.4$ arcsecond from the tracking center in Right Ascension and $\Delta \delta=-1.2$ arcsecond in Declination.  
The {map nodes} of other channels were almost the same as that of $V_{\rm LSR}=-32.31$ km s$^{-1}$, within the coarse resolution. 
The fringe rate map did not change significantly throughout observations. 

Next we made dirty map data cubes using the AIPS task IMAGR to search for maser spots within a field of $\timeform{1''.25} \times \timeform{1''.25}$ around the position determined with the fringe rate map, $(\Delta \alpha \cos{\delta}, \Delta \delta)=(\timeform{-0''.4}, \timeform{-1''.2})$. We searched a velocity range of $V_{\rm LSR}=-56.33$ {to} $-23.04$ km s$^{-1}$, to cover all the emitting components found in the range $V_{\rm LSR}=-50.43$ {to} $-32.31$ km s$^{-1}$. 
We cataloged maser spots found in each observation track and performed the CLEAN procedure \citep{hog74} for each maser spot using IMAGR. Finally, the position of each spot was measured using the AIPS task JMFIT. 
%When peak position 

\section{Results}
\subsection{Detected Maser Features}
{Maser features were identified by the following criteria: the emission must be detected at a level of greater than five times the rms noise in individual channel maps and must appear {over} adjacent spectral channels.} %Since the the position of the peak moved slightly channel by channel, the position of a maser feature was defined as the midpoint of the maximum and minimum values in each axis. The error was defined as the difference between the midpoint and the maximum or minimum value plus the position accuracy (half beam size divided by signal-to-noise ratio).   
All detected maser features are listed in table \ref{maser-list}, {showing:} (1) maser feature ID, (2) initial R.A. offset ($\Delta \alpha_0 \cos{\delta}$), (3) R.A. proper motion ($\mu_\alpha \cos{\delta}$), (4) initial Decl offset ($\Delta \delta_0$), (5) R.A. proper motion ($\mu_\delta$), (6) LSR velocity ($V_{\rm LSR}$), and (7) Detection epochs. 

The center of the maser distribution map was redefined as $(\alpha_{\rm c}, \delta_{\rm c})=(\timeform{21h39m40s.501608}, \timeform{+51D20'32''.60108})$ (J2000.0), which was the averaged position of the all detected maser features in the first epoch.  

Typically, more than 10 features were detected in most epochs, with the exceptions of epochs 5 and 8. In total, 39 maser features were detected. Five features were identified in six or more epochs and another 13 features were detected in two two four epochs. The other 21 features were detected only once. 
The identification criteria were (1) an identical maser feature moves less than 2.5 mas from its position in the previous observation, (2) a difference in velocity giving peak temperature between consecutive observations is less than 1 km s$^{-1}$, and (3) R.A. or Decl offsets must meet the theoretical curve of the annual parallax and linear proper motion within an error of 1 mas. The first criterion {excludes large proper motions exceeding 10 mas yr$^{-1}$ considering a typical observation interval of 90 days}. 
%The general distribution of all maser features is shown in figure 5. %\ref{spotmap}. 

\subsection{Annual Parallax and Proper Motion}
We used the five features (ID 1--5 of table \ref{maser-list}) detected in six or more epochs to calculate the annual parallax. Annual parallax fitting was {performed} for each maser spot. {The annual} parallax was calculated {using} 12 maser spots {associated with 5 individual maser features.} {In addition, we performed a combined-fitting in which the astrometry of all maser spots are fit together assuming a common distance (see for example \citet{sak12}). The combined-fitting produced a result which is of higher accuracy and consistent with that obtained for individual spots.}
%In addition, we conducted the {combined-fitting, which is annual parallax fitting made for all maser spots simultaneously assuming the common parallax as applied in former works \citep{sak12}. As a result, the annual prallax was obtained with the better accuracy by the combined-fitting and are} consistent with the result obtained with each maser spot. 
The results are shown in table \ref{parallax_spot}. 
Figure 4 shows the R.A. and Decl offsets of maser features plotted against time. 
The time variation of the R.A. offset $\Delta \alpha \cos{\delta}^{(i)}$ and Decl offset $\Delta \delta^{(i)}$ due to annual parallax and linear proper motion of $i$th feature is expressed as a function of time $t$, as follows.
\begin{eqnarray}
 \Delta \alpha \cos{\delta_{\rm c}}^{(i)} &=& \varpi (-\sin{\alpha_{\rm c}}\cos{\odot}+\cos{\varepsilon}\cos{\alpha_{\rm c}}\sin{\odot})\\
                                  & &+\mu_\alpha^{(i)} \cos{\delta_{\rm c}} (t-t_0^{(i)}) + \Delta\alpha_0^{(i)}\cos{\delta_{\rm c}}\nonumber\\
 \Delta \delta^{(i)}              &=& \varpi (-\cos{\alpha_{\rm c}}\cos{\odot}+\cos{\varepsilon}\sin{\alpha_{\rm c}}\sin{\odot})\sin{\delta_{\rm c}}\nonumber\\
                                  & &+\sin{\varepsilon}\cos{\delta_{\rm c}}\sin{\odot}\nonumber\\
                                  & &+\mu_\delta^{(i)} (t-t_0^{(i)}) + \Delta\delta_0^{(i)}
\end{eqnarray}
where $\varpi$ is the anuall parallax, $\varepsilon$ {is} the obliquity ($\varepsilon=23.440972$), $\odot$ is the ecliptic longitude of the Sun, $\mu_\alpha^{(i)}$ and $\mu_\delta^{(i)}$ are proper motion in the R.A. and Decl directions, $t_0^{(i)}$ is the first detection time of the $i$th feature, and $\Delta \alpha_0\cos{\delta}^{(i)}$ and $\Delta \delta_0^{(i)}$ are R.A. and Decl offsets at the time $t_0$. 
We fitted this theoretical curve to the observed positions of the detected features by the least squares method, {which is explained in \citet{kov12}}, with known parameters $\Delta \alpha \cos{\delta_{\rm c}}^{(i)}$, $\Delta \delta^{(i)}$, $\alpha_{\rm c}$, $\delta_{\rm c}$, $\odot$, $\varepsilon$, $t$, and $t_0^{(i)}$, to obtain unknown parameters $\varpi$, $\mu_\alpha\cos{\delta_{\rm c}}^{(i)}$, $\mu_\delta^{(i)}$, $\Delta \alpha_0\cos{\delta_{\rm c}}^{(i)}$ and $\Delta \delta_0^{(i)}$. As seen in figure \ref{Parallax}, 
most of the data were well fitted by the theoretical curve. 
We adopted {positional} errors of 0.24 mas, {which were required to produce a reduced chi squared of one}, and {are} indicated as the error bars in figure \ref{Parallax}. 
%Motions of the features were well fitted with a theoretical curve of annual parallax and linear proper motion. 

The obtained annual parallax was $0.262 \pm 0.031$ mas, which corresponds to Heliocentric distance of $D=3.82^{+0.51}_{-0.41}$ kpc. Hereafter, we adopt this value as the distance of IRAS 21379+5106. 

The averaged proper motions of 12 maser spots are also summarized in table \ref{maser-list}. 
The proper motion of this system was calculated to be $\mu_\alpha \cos{\delta}=-2.74 \pm 0.08$ mas yr$^{-1}$,  $\mu_\delta=-2.87 \pm 0.18$ mas yr$^{-1}$, which was the averaged value of the five maser features, adopting a standard error. 
%The reason that the other masers (ID 6 and 7) were not included in the calculation of the systemic proper motion was that they have excessively large internal motion, which is not appropriate for this purpose, as mentioned at the end of this subsection. 
%, where the internal motion was estimated by subtracting the averaged systemic proper motion. 
The systemic proper motion in {Galactic coordinates} was calculated to be $\mu_l \cos{b}=-3.95\pm 0.18$ mas yr$^{-1}$ and $\mu_b=-0.34\pm 0.08$ mas yr$^{-1}$. Note that the error vector $(\Delta \mu_l \cos{b}, \Delta \mu_b)=(0.18, 0.08)$ mas yr$^{-1}$ in the Galactic coordinate was calculated with the same coordinate transformation as the error vector $(\Delta \mu_\alpha \cos{\delta}, \Delta \mu_\delta)=(0.08, 0.18)$ mas yr$^{-1}$ in the equatorial coordinate. 
The corresponding linear velocity was $V_l=D \mu_l \cos{b}=-71.5 \pm 10.1$ km s$^{-1}$ and $V_b=D\mu_b=-6.21\pm 1.70$ km s$^{-1}$ in Galactic coordinates, when using $D=3.82$ kpc. 

We also estimated proper motions of the other 13 maser features detected in only two to four epochs, in addition to the {other} five features, adopting the newly determined systemic distance of $D=3.82$ kpc. 

Table \ref{internal_motion} shows the internal motion of IDs 1--18, which were calculated by subtracting the systemic proper motion, $(\overline{{\mu_\alpha}\cos{\delta}},\overline{{\mu_\delta}})=(-2.74\pm0.08, -2.87\pm 0.18)$ mas yr$^{-1}$ from the individual proper motions $({{\mu_\alpha}\cos{\delta}},{{\mu_\delta}})$, and by subtracting the systemic $V_{\rm LSR}=-42.27\pm 0.22$ km s$^{-1}$ from those of the individual masers. The result is discussed in section 5.4. 

%All proper motions are listed in Table \ref{internal_motion}, and vector map is also shown in figure \ref{vectmap}. %Note that a maser with too large proper motion (ID:6) and masers with large proper motion error more than 100 mas yr$^{-1}$ (ID:15 and 16) were not included in this figure and were not included in the discussion hereafter.  

\subsection{3-D Motion Relative to the Sun in the LSR coordinate}
We calculated the three dimensional velocity {vectors} $(U, V, W)$ with respect to the LSR, using the LSR velocity, obtained parallax, and proper motion, following \citet{rei09}. As shown in figure 6, 
%\ref{UV_VRVtheta}, 
we defined the LSR velocity coordinate so that $U$, $V$, and $W$ axes were aligned with $(l,b)=(0, 0)$, $(90^\circ, 0)$, and $(0, 90^\circ)$, respectively, %following the recently well used notation \citep{bin98}, 
though the $U$ axis is opposite to the classical definition \citep{mih81}.  

%\subsubsection{Heliocentric Velcocity}
First, the LSR velocity $V_{\rm LSR}=-42.27\pm 0.22$ km s$^{-1}$ was converted to the heliocentric velocity $V_{\rm Hel}=-56.43 \pm 0.22$ km s$^{-1}$ with 
\begin{equation}
 V_{\rm Hel} = V_{\rm LSR} - (U_\odot^{\rm Std} \cos{l}+V_\odot^{\rm Std} \sin{l} )\cos{b} - W_\odot^{\rm Std} \sin{b}
\end{equation}
where $(U_\odot^{\rm Std}, V_\odot^{\rm Std}, W_\odot^{\rm Std})$ is the standard solar motion defined as $U_\odot^{\rm Std}=10.3$ km s$^{-1}$, $V_\odot^{\rm Std}=15.3$ km s$^{-1}$, and $W_\odot^{\rm Std}=7.7$ km s$^{-1}$, whose absolute value and direction are 20 km s$^{-1}$ and $\alpha(1900)=18^{\rm h}$, $\delta(1900)=+30^{\rm d}$, respectively. 

%\subsubsection{3-D Motion relative to the Sun}
Second, the linear velocity vector $(V_l, V_b)$ was converted to a 3-D vector, relative to the Sun, as follows. 

\begin{eqnarray}
U-U_\odot     &=& (V_{\rm Hel}\cos{b}-V_b\sin{b})\cos{l} -V_l\sin{l}\\
V-V_\odot     &=& (V_{\rm Hel}\cos{b}-V_b\sin{b})\sin{l} +V_l\cos{l}\\
W-W_\odot     &=& V_b\cos{b} + V_{\rm Hel}\sin{b}
\end{eqnarray}
The obtained 3-D velocity vector relative to the Sun in the LSR coordinate was $U-U_\odot=76.4\pm 10.4$ km s$^{-1}$, $V-V_\odot=-49.7\pm 1.0$ km s$^{-1}$, and $W-W_\odot=-5.29\pm 1.70$ km s$^{-1}$.

\section{Discussion}
\subsection{Absolute 3-D Motion in the Galaxy and the Rotation Velocity}
We first discuss the 3-D motion of the maser source, IRAS 21379+5106, in the Galaxy, based on the obtained values of the parallax and proper motion. 

\subsubsection{3-D Vector in the LSR coordinate}
The 3-D motion of the object in the LSR coordinate was calculated by adding the solar peculiar motion $(U_\odot, V_\odot, W_\odot)$ to the relative velocity obtained in the previous section. 

Thanks to the recent improvements that come with the continuous developments in observational techniques and methods, the solar peculiar motion relative to the LSR has been measured more accurately than the classical value of $(U_\odot^{\rm Std}, V_\odot^{\rm Std}, W_\odot^{\rm Std})$. \citet{deh98b} estimated it to be $U_\odot=10.0$ km s$^{-1}$, $V_\odot=5.2$ km s$^{-1}$, and $W_\odot=7.2$ km s$^{-1}$ based on the parallax and proper motions obtained from the Hipparcos catalog \citep{per97}. \citet{sch10} estimated the solar 3-D motion of the LSR so that it should be insensitive to the metallicity gradient and concluded it to be $U_\odot=11.1$ km s$^{-1}$, $V_\odot=12.2$ km s$^{-1}$, and $W_\odot=7.3$ km s$^{-1}$. \citet{hon12} took the same values for $U_\odot$ and $V_\odot$ as \citet{deh98b} and measured $V_\odot$ to be 12.0 km s$^{-1}$ by analyzing VLBI astrometry data for 52 Galactic maser sources. Comparing these 3 papers with each other, it can be regarded that $U_\odot$ and $W_\odot$ are well determined. Regarding the other component $V_\odot$, the results from \citet{sch10} and \citet{hon12} are consistent with each other. Therefore, we adopt the solar peculiar motion relative to the LSR as $U_\odot=10.0$ km s$^{-1}$, $V_\odot=12.0$ km s$^{-1}$, and $W_\odot=7.2$ km s$^{-1}$ as adopted by \citet{hon12}. We conclude that the 3-D velocity vector of the IRAS 21379+5106 in the LSR coordinate is $U=86.4\pm 10.1$ km s$^{-1}$, $V=-37.7\pm 1.0$ km s$^{-1}$, and $W=1.91\pm 1.70$ km s$^{-1}$. 

\subsubsection{Absolute 3-D Motion in the Galaxy}
With the 3-D vector of the source in the LSR coordinate, we can estimate its absolute 3-D motion in the Galaxy. 

First, the Galactocentric distance to the source is calculated using the following equation,  
\begin{equation}
 R=\sqrt{R_0^2+(D\cos{b})^2-2R_0D\cos{b}\cos{l}}. 
\end{equation}

Next we define the velocity vector as $(V_{R_{\rm in}}, V_\theta, V_z)$, where $V_{R_{\rm in}}, V_\theta, V_z$ are the velocity components directed to the Galactic center, the rotational velocity about the Galactic center, and the vertical velocity perpendicular to the Galactic plane (the direction of the north Galactic pole is positive), respectively, as outlined in figure 6. %\ref{UV_VRVtheta}. 
Using this definition, each component is estimated using the following equations,
\begin{eqnarray}
 V_{R_{\rm in}}&=&U\cos{\phi} - (V+V_0)\sin{\phi}\\
 V_\theta      &=&(V+V_0)\cos{\phi} + U\sin{\phi}\\
 V_z           &=& W
\end{eqnarray}
where $\phi$ is the angle between the Sun and the source as viewed from the Galactic center given by 
\begin{equation}
 \sin{\phi}={D\cos{b}\over R}\sin{l}\\
 \cos{\phi}={R_0-D\cos{b}\cos{l}\over R}. 
\end{equation}

Since the velocity vector $(V_{R_{\rm in}}, V_\theta, V_z)$ depends on the Galactic constants, 
$R_0$ and $V_0$, which are the Galactocentric distance of the Sun and the rotational velocity of the Sun about the Galactic center, respectively, we estimated the 3-D velocity vector using some possible values reported in recent studies. 

First we took the Galactic constants from \citet{hon12}, who concluded that the Galactic constants are $R_0=8.05$ kpc and $V_0=238$ km s$^{-1}$ by analyzing the available latest astrometric data obtained with VLBI including VERA, VLBA, and EVN. Using these Galactic constants we estimate the Galactocentric distance of IRAS 21379+5106 to be $R=9.22$ kpc and the 3-D velocity components, $V_{R_{\rm in}}=-3.8$ km s$^{-1}$, $V_\theta=218$ km s$^{-1}$, and $V_z=1.91$ km s$^{-1}$.

If we take IAU recommendation value, $R_0=8.5$ kpc and $V_0=220$ km s$^{-1}$ \citep{ker86}, which are the most commonly adopted values, we estimate a Galactocentric distance of $R=9.63$ kpc and the velocity vector components $V_{R_{\rm in}}=7.5$ km s$^{-1}$, $V_\theta=202$ km s$^{-1}$, and $V_z=1.91$ km s$^{-1}$. 
  
According to \citet{deh98a}, a mass model with $R_0=8.0$ kpc and $V_0=217$ km s$^{-1}$ is best fitted to the observational data. Adopting these Galactic constants, we estimate a Galactocentric distance of $R=9.18$ kpc and velocity vector components $V_{R_{\rm in}}=4.4$ km s$^{-1}$, $V_\theta=199$ km s$^{-1}$, and $V_z=1.91$ km s$^{-1}$. 

Finally when we adopt the Galactic constants $R_0=8.34$ kpc and $V_0=240$ km s$^{-1}$, recently estimated by \citet{rei14} by analyzing astrometry data with VLBA, the Galactocentric distance and the velocity vectors are estimated as $R=9.48$ kpc, $V_{R_{\rm in}}=-1.9$ km s$^{-1}$, $V_\theta=220$ km s$^{-1}$, and $V_z=1.91$ km s$^{-1}$, respectively. These results are summarized in table \ref{3D_velocity}. 

%Firstly, t

These results systematically show that the rotational velocity $V_\theta$ tends to be slower than that of the Sun. %, though the error is not small enough. 
This is consistent with the declining rotation curve beyond $R=R_0$ obtained with the recent work on maser sources associated with the Perseus arm \citep{sak12,oh10} and is consistent with the existence of 9-kpc dip found in the Galactic rotation curve \citep{sof09}, which might be caused by the mass excess at the Galactocentric distance of 11 kpc. Regarding the radial velocity component, $V_{R_{\rm in}}$, the error was not small enough to judge which direction the object moves. 
The vertical component $V_z=1.91$ km s$^{-1}$ is small enough to be consistent with the natural random velocity, whose typical velocity dispersion is 2 -- 11 km s$^{-1}$ \citep{mal94}.

\subsection{Spectral Type of YSO}
Our accurately measured distance modifies the bolometric luminosity of the central young stellar object (YSO). \citet{wan09} estimated it to be $5.18 \times 10^3 L_\odot$, which corresponds to the spectral type B1 based on IRAS data and a kinematic distance of $5.9$ kpc, when it is assumed to be a Zero Age Main Sequence star (ZAMS)\citep{pan73}. 
With the newly determined distance, the bolometric luminosity was modified to be $(2.15\pm 0.54)\times 10^3 L_\odot$ by multiplying with the square of the distance ratio $(3.8/5.9)^2$. As a result, we re-estimate the spectral type of the central YSO to be B2--B3. This result is still consistent with the fact that this object is known to have an ultra compact H {\sc ii} region, such objects are associated with YSOs of main sequence luminosity corresponding to type B3, or hotter. 
This picture is valid even though the object was found to be less luminous than previously thought. 

\subsection{Distance to the Perseus Arm}
Since IRAS 21379+5106 is a member of the Perseus arm of the Milky Way Galaxy, we also {consider} that its trigonometric distance can be regarded as that of the Perseus arm.  
%$3.76^{+0.81}_{-0.56}$ kpc, in the direction of $(l,b)=(95\fdg29635672, -0\fdg93695671)$.
{We} compare our result with other {relevant distance} values. 

{\noindent{\it 1. Another VLBI astrometric measurement}: {The annual} parallax of IRAS 21379+5106 was recently reported by \citet{cho14}, {who reported an} annual parallax {of} $0.206 \pm 0.007$ mas and {a} corresponding trigonometric distance {of} $4.85^{+0.17}_{-0.16}$ kpc, based on {astrometric observations with the VLBA.}
{The difference in parallax values} is about $2\sigma$. {A} possible reason for this discrepancy is that they used only {a} single feature {rather than the cumulative results of multiple maser features, as was done} in this paper. %Our resultant trigonometric distance is consistent with those of other sources associated with Perseus arm near IRAS 21379+5106 such as G094.60−1.79 and G100.37−3.57 reported in \citet{cho14}. }  

\noindent{\it 2. Kinematic Distance}: The kinematic distance is often used as a distance estimation and \citet{wan09} calculated it to be 5.9 kpc. If we {assume a} flat rotation curve, 
% in order to consider the case other Galactic constants $R_0$, and $V_0$, 
the Galactocentric distance $R$ and the Heliocentric distance $D$ are calculated using the LSR velocity as follows. 
%\begin{equation}
%R = R_0 \left({V_0 \over V_{\rm LSR}}{1 \over \sin{l}\cos{b}} + 1 \right)^{-1} \\
%D = R_0 \cos{l} + \sqrt{R^2 - R_0^2\sin^2{l}}
%\end{equation}
\begin{eqnarray}
R &=& R_0 \left({V_0 \over V_{\rm LSR}}{1 \over \sin{l}\cos{b}} + 1 \right)^{-1} \\
D &=& R_0 \cos{l} + \sqrt{R^2 - R_0^2\sin^2{l}}
\end{eqnarray}
When we adopted Galactic constants we use in this paper, $R_0=8.05\pm 0.45$ kpc and $V_0=238\pm 14$ km s$^{-1}$, we have $R=9.78\pm 0.56$ kpc and $D=4.86\pm 1.30$ kpc, which turn out be overestimations, compared to our trigonometric distance. Adopting any of the other Galactic constants mentioned in the previous section, the kinematic distance is still larger than the trigonometric distance, as long as we assume a flat rotation curve. 

%When we used the rotation curve derived by \citet{deh98a} and adopting the Galactic constants $R_0=8.0$ kpc and $V_0=217$ km s$^{-1}$, the Galactocentric distance is calculated to be $R=9.79$ kpc and the Heliocentric distance $D=4.96$ kpc, which is almost the same as the case of the flat rotation curve. 
 
\noindent{\it 3. S124}: Sharpless 124 (S124) is an H {\sc ii} region seen in figure 1 
%\ref{fig:RGBmap} 
as a circular-shaped emission region just below IRAS 21379+5106. With celestial coordinates and LSR velocity $(l,b)=(94\fdg57, -1\fdg45)$ and $V_{\rm LSR}=-43.4$ km s$^{-1}$, it is closely associated with IRAS 21379+5106, thus these regions are thought to coexist. 
%S124 appears somehow connected to the IRAS21379+5106. 
\citet{bra93} estimated its distance to be $2.6 \pm 0.6$ kpc by photometric distance, which underestimated it by 30\%. 
However, our newly determined trigonometric distance seems more plausible 
because applying the photometric distance of S124 to IRAS 21379+5106 would re-classify the central YSO as a spectral type later than B3, which would not be hot enough to produce the observed ultra compact H {\sc ii} region.

\noindent{\it 4. AFGL 2789}: AFGL 2789 is another object located in the Perseus arm, and its position $(l,b)=(94\fdg6, -1\fdg80)$ and LSR velocity ($-44$ km s$^{-1}$) are close to those of IRAS 21379+5106 \citep{oh10}. This object is also seen in figure 1 
%\ref{fig:RGBmap} 
as a bright point source located at the southern edge of the circular-shaped infrared emission of S124. {\citet{oh10} concluded that} trigonometric distance of AFGL 2789 was $3.08 \pm 0.30$ kpc, which is consistent with our study, within the error.

\subsection{Asymmetric Spectrum and Maser Distribution --- Evidence of Bipolar Outflow}
The maser distribution shown in figure 5 indicates that most of the maser features seem to be linearly aligned on the sky plane. North-east and south-west components tend to be red- and blue-shifted, respectively. We drew a straight line in figure 5, 
%\ref{spotmap}, 
which is fitted with the positions of all the maser features listed in table \ref{internal_motion} except IDs 3, 4, 8, and 9 which {deviate} largely from the line. The observed situation resembles a bipolar outflow from the YSO, which is reasonable considering this object is an ultra-compact H {\sc ii} region and high-mass star-forming region \citep{bro96,wan09,wu06,she96}, which are often associated with molecular outflows \citep{ort12,qin08,tof95}. 

The blue-shifted component is dominant in the maser distribution map. This corresponds to the asymmetry of the spectrum shown in figure 2, 
%\ref{CPSPEC}, 
where the blue component is brighter than the red one. This asymmetry is also found in the $^{12}$CO($J=$1--0) and $^{13}$CO($J=$3--2) spectra obtained with the IRAM and KOSMA telescopes, respectively \citep{wou89,wan09}. 

A plausible explanation of this characteristics is that there exists a disk around the central young star and the red component is obscured by this optically thick disk, as seen in \citet{mot13}. 

Next let us consider the internal motions of the maser features shown in table \ref{internal_motion} and figure 7. %\ref{vectmap}. 
Maser features seem to be ejected from the central YSO, which is thought to be located on the fitted line. These are the similar kinematic characteristics as were found in the proper motion of water-vapor masers associated with IRAS 21391+5801 \citep{pat00}. 
In a further detail, the directions of vectors are not always aligned with the fitted line. This implies that the motions of the maser features are not simple ballistic motions from the central YSO, but seem perturbed possibly due to interaction with ambient matter \citep{lad96,yu99} and/or helical motion driven by a magnetic field \citep{mat99,woi05}.

The maser features located in the region of Decl offset $-100$ mas to $-150$ mas seem to move eastwards, differently from the bipolar flow motion traced by the other maser features. A possible explanation for this discontinuity is that the bipolar flow hits against the ambient interstellar medium {which} subsequently changed the direction of {the} outflow.

\section{Summary}
We measured the annual parallax and proper motion of the high-mass star-forming region IRAS 21379+5106 with VERA. The resultant annual parallax was $0.262 \pm 0.031$ mas, which corresponds to a trigonometric distance of $3.82^{+0.51}_{-0.41}$ kpc. The proper motion was $\mu_\alpha \cos{\delta}=-2.74 \pm 0.08$ mas yr$^{-1}$ and $\mu_\delta=-2.87 \pm 0.18$ mas yr$^{-1}$. 
%The three-dimensional velocity vector in the LSR coordinate was estimated to be $U=87.5\pm 10.4$ km s$^{-1}$, $V=-37.6\pm 1.0$ km s$^{-1}$, and $W=2.50\pm 2.10$ km s$^{-1}$. 
When we adopt the Galactic constants $R_0=8.05\pm 0.45$ kpc and $V_0=238\pm 14$ km s$^{-1}$, suggested by \citet{hon12}, we conclude that the Galactocentric distance of IRAS 21379+5106 is $R=9.22\pm 0.43$ kpc and the velocity components in cylindrical coordinates are $V_{R_{\rm in}}=-3.8\pm 16.9$ km s$^{-1}$, $V_\theta=218\pm 19$ km s$^{-1}$, and $V_z=1.91\pm 1.70$ km s$^{-1}$. 
Using our new distance, the spectral type of the central YSO was re-estimated to be B2--B3, but not B1, which was its previous classification. {Masers are distributed along a} straight line {indicating} the existence of a bipolar outflow.  
All the derived physical parameters are summarized in table \ref{summary}.

We thank the VERA project members for  supporting observations. We would also like to thank the referee for carefully reading the manuscript.  The authors  acknowledge financial support from JSPS/MEXT KAKENHI Grant Numbers 26800104, 13J03569, 25610043, 24540242, and 25120007.

\begin{figure}[h]
  \begin{center}
    \FigureFile(80mm,80mm){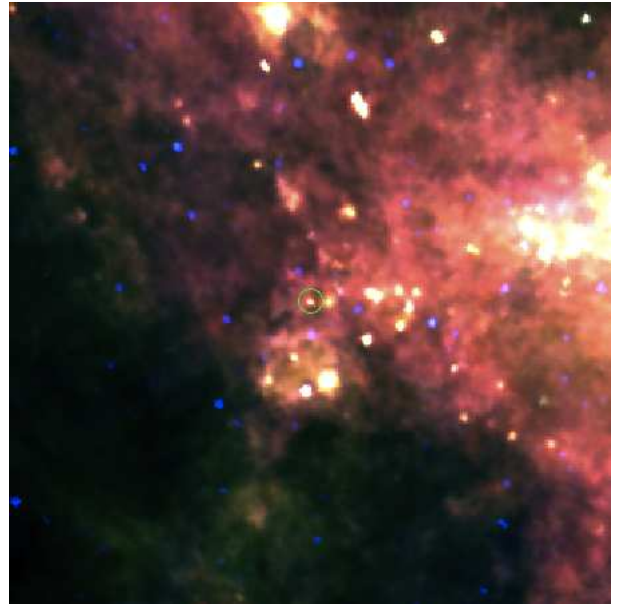}
    %%% \FigureFile(width,height){filename}
  \end{center}
  \caption{Mid-Far infrared {composite} image of $7\fdg5\times 7\fdg5$ area {centered} on IRAS 21379+5106. This was made using IRIS 25 $\mu$m-, 60 $\mu$m-, and 100 $\mu$m-band images. IRAS 21379+5106 is indicated with a green circle.  A circular-shaped emission region just below IRAS 21379+5106 is {the} H {\sc ii} region Sharpless 124 (S124). A bright point source located at the southern edge of the circular-shaped infrared emission of S124 is AFGL 2789. }\label{fig:RGBmap}
\end{figure}

\begin{figure}[h]
  \begin{center}
    \rotatebox{-90}{\FigureFile(60mm,80mm){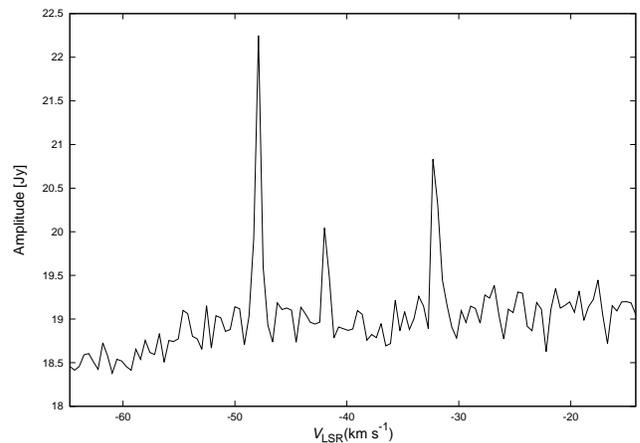}}
    %%% \FigureFile(width,height){filename}
  \end{center}
  \caption{Cross-power spectrum obtained at the baseline between Mizusawa and Iriki at UT 7:39:47.3 of 6th November 2009. The channel range is from 140 ($V_{\rm LSR}=-14.19$ km s$^{-1}$)to 260 ($V_{\rm LSR}=-64.75$ km s$^{-1}$). }\label{CPSPEC}
\end{figure}

\begin{figure}
%  \begin{center}
    \rotatebox{-90}{\FigureFile(80mm,80mm){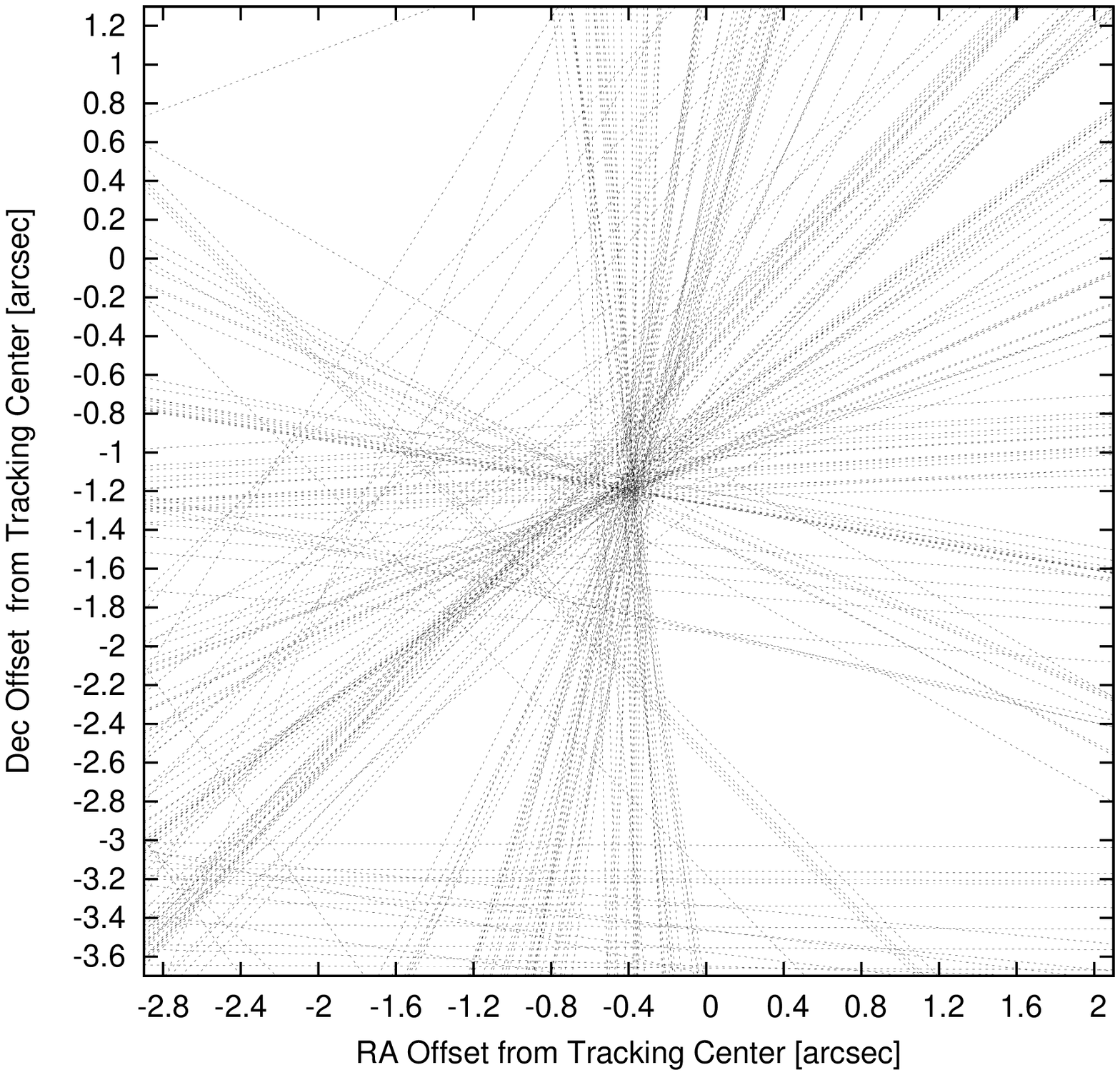}}
    %%% \FigureFile(width,height){filename}
\caption{Fringe-rate map of the maser emission detected at the channel of $V_{\rm LSR}=-32.31$ km s$^{-1}$ in the first epoch. The map origin is located at the phase tracking center $(\alpha, \delta)=(21:39:40.55, +51:20:34.0)$ (J2000.0). The position of the maser spot is shifted by $\Delta \alpha\cos{\delta}=-0.4$ mas from the phase tracking center in Right Ascension and $\Delta \delta=-1.2$ mas in Declination. }\label{FRM}
%  \end{center}
\end{figure}

\begin{figure*}
%  \begin{center}
\FigureFile(85mm,200mm){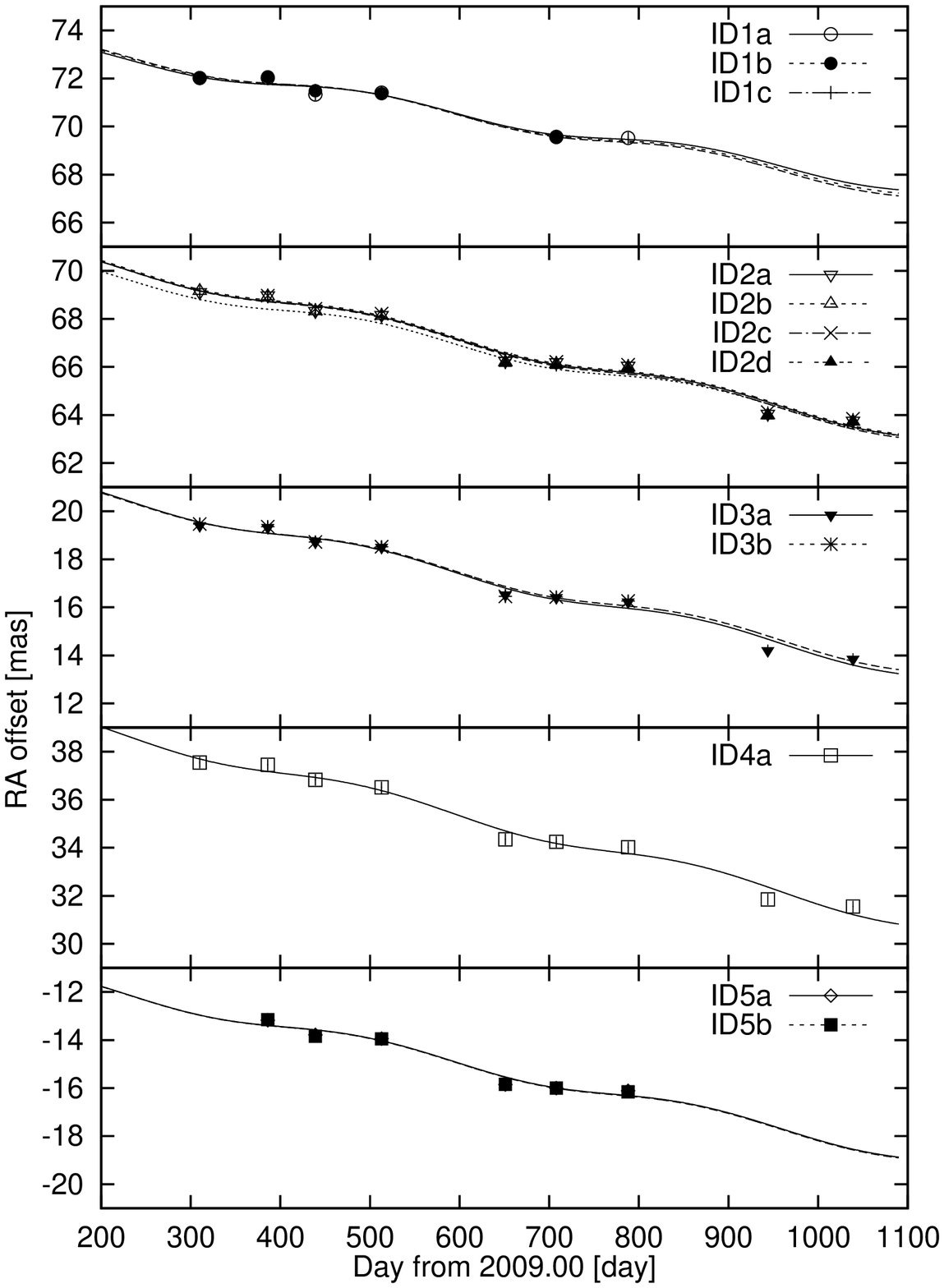}
\FigureFile(85mm,200mm){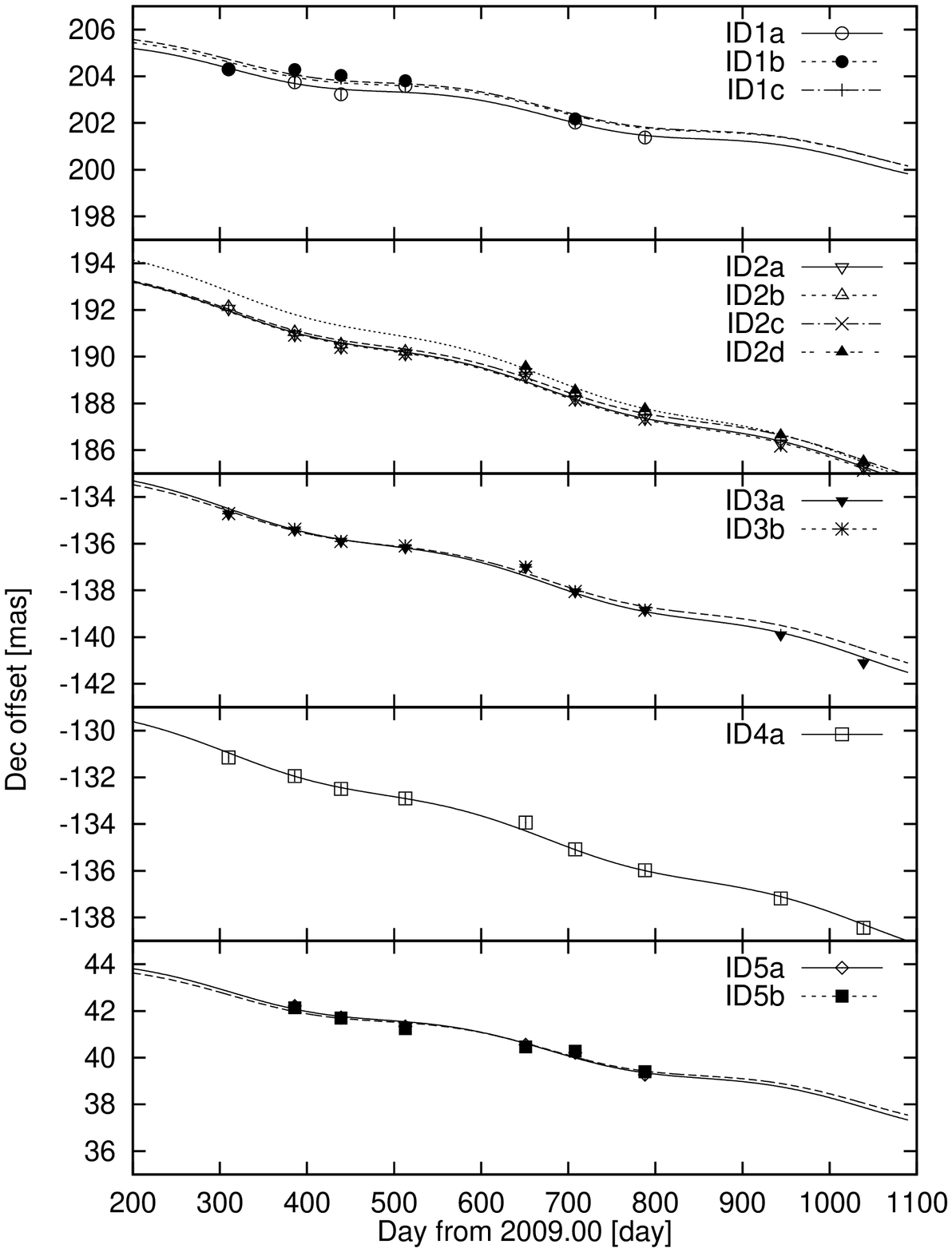}
\FigureFile(85mm,85mm){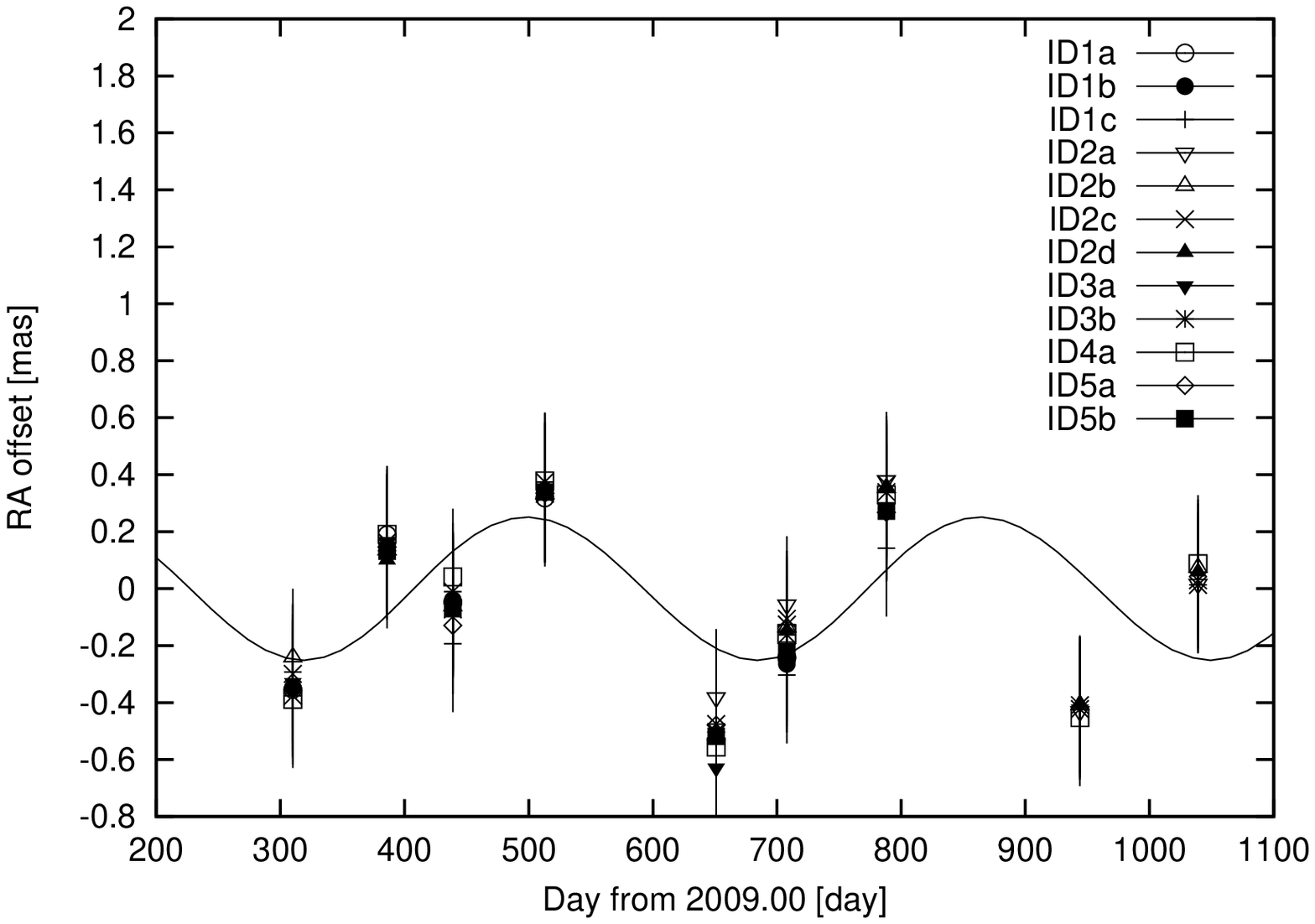}
\FigureFile(85mm,85mm){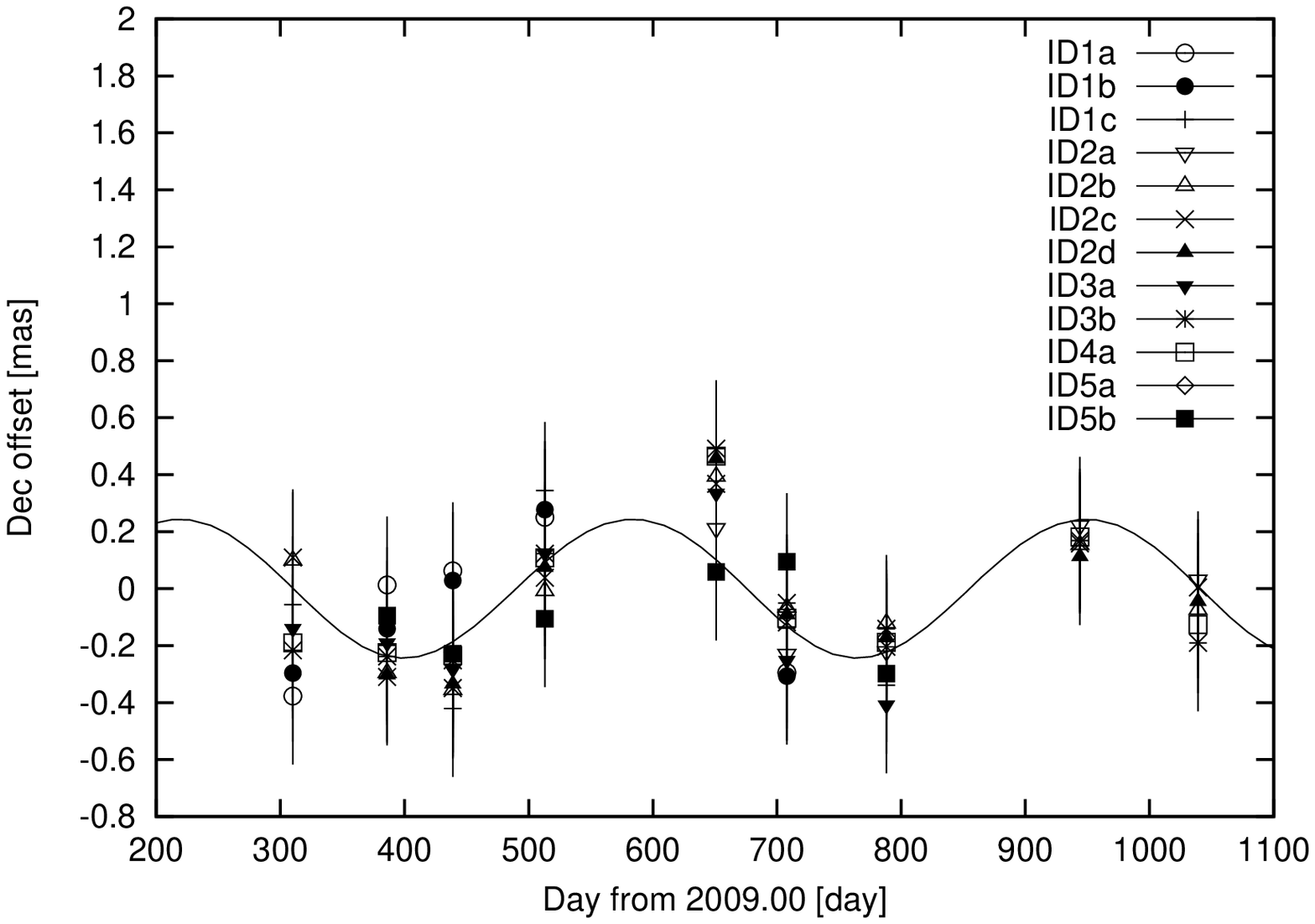}
    %%% \FigureFile(width,height){filename}
\caption{Top: Time variations of R.A. (left) and Decl (right) offsets. The potted curves are the best-fitted models including annual parallax and linear proper motion. Bottom: Time variation of R.A. (left) and Decl (right) offsets with the proper motions subtracted. }\label{Parallax}
%  \end{center}
\end{figure*}

\begin{figure*}
%  \begin{center}
    \FigureFile(100mm,100mm){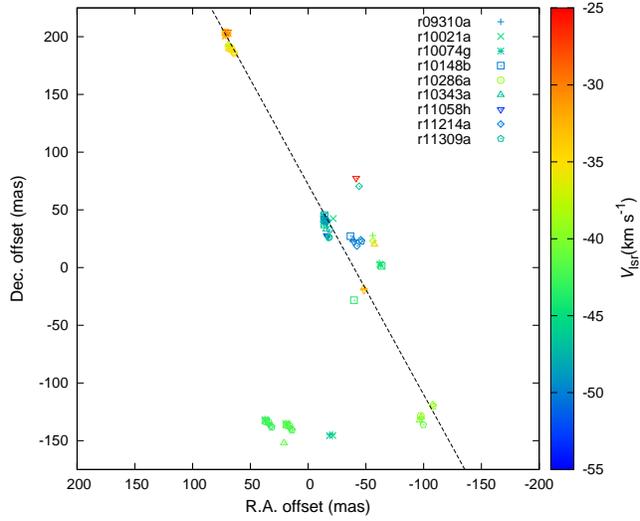}
    %%% \FigureFile(width,height){filename}
\caption{Map of maser features listed in table 2. The color denotes the line-of-sight velocity and its value is indicated in the vertical color bar. Each symbol denotes the observation epoch shown at the top-right corner. Each feature shows movement between epochs due to the proper motion and annual parallax. {The angular size of 1 mas corresponds to a physical size of 3.82 AU. }}\label{spotmap}
%  \end{center}
\end{figure*}

\begin{figure*}
  \begin{center}
    \FigureFile(100mm,100mm){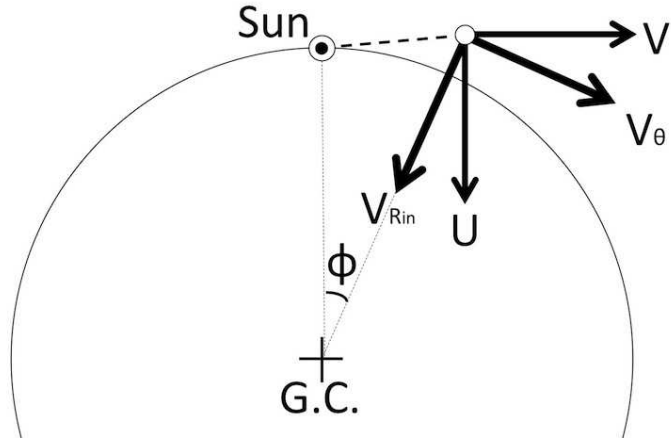}
    %%% \FigureFile(width,height){filename}
\caption{Schematic depiction {definining} the velocity vectors $(U,V)$ and $(V_{R_{\rm in}},V_\theta)$. }\label{UV_VRVtheta}
  \end{center}
\end{figure*}

\begin{figure*}
%  \begin{center}
    \FigureFile(100mm,100mm){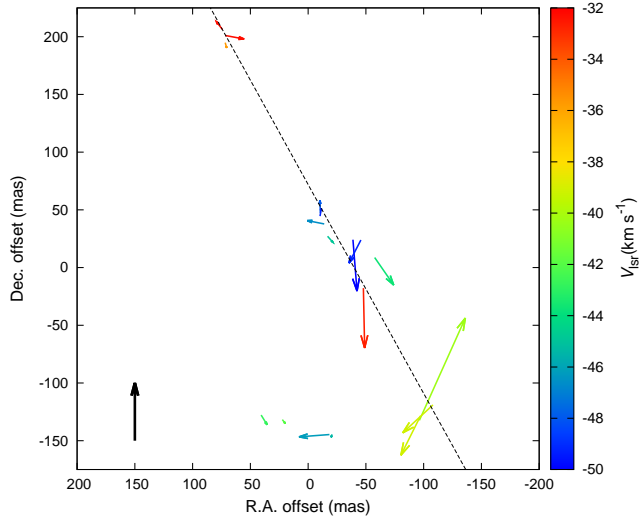}
    %%% \FigureFile(width,height){filename}
\caption{Vector map of the internal motion made with the proper {motions} listed in table \ref{internal_motion}. The internal motion is obtained by subtracting the averaged motion {individual feature motions.} The black thick arrow at the bottom-left indicates vector of 100 km s$^{-1}$. {The angular size of 1 mas corresponds to a physical size of 3.82 AU. }}\label{vectmap}
%  \end{center}
\end{figure*}

\begin{table*}
  \caption{Summary of Observations}\label{obssummary}
  \begin{center}
    \begin{tabular}{rrrrrr}
      \hline
      \hline
Epoch & Date        & DOY$^\dagger$ & Synthesized beam  & PA       & Noise            \\
      &             &     &(mas $\times$ mas) & ($\deg$) & (mJy beam$^{-1}$)\\
      \hline
1     & 2009 NOV 06 & 310 & 1.21 $\times$ 0.75 & $-58$    & 7.3              \\
2     & 2010 JAN 21 & 386 & 1.18 $\times$ 0.75 & $-48$    & 8.7              \\
3     & 2010 MAR 15 & 439 & 1.20 $\times$ 0.77 & $-60$    &13.3              \\
4     & 2010 MAY 28 & 513 & 1.16 $\times$ 0.77 & $-48$    & 9.7              \\
5     & 2010 OCT 13 & 651 & 1.21 $\times$ 0.73 & $-41$    & 3.8              \\
6     & 2010 DEC 09 & 708 & 1.21 $\times$ 0.74 & $-41$    & 4.2              \\
7     & 2011 FEB 27 & 788 & 1.13 $\times$ 0.73 & $-48$    & 6.3              \\
8     & 2011 AUG 02 & 944 & 1.14 $\times$ 0.74 & $-43$    &13.2              \\
9     & 2011 NOV 05 &1039 & 1.14 $\times$ 0.73 & $-39$    &14.5              \\
      \hline
     \hline
    \end{tabular}

$^\dagger$ DOY (Day-of-year) indicates days from 2009 January 1. 
  \end{center}
\end{table*}

%\begin{sidewaystable}
\begin{longtable}[width=450pt]{rrrrrrr}
  \caption{List of Maser features}\label{maser-list}
  \hline          
\multicolumn{1}{r}{ID}  &\multicolumn{1}{c}{$\Delta \alpha_0\cos{\delta}$}  & \multicolumn{1}{c}{$\mu_\alpha\cos{\delta}$} & \multicolumn{1}{c}{$\Delta \delta_0$}          & \multicolumn{1}{c}{$\mu_\delta$}      & \multicolumn{1}{c}{$V_{\rm LSR}$}   & Detection \\
\multicolumn{1}{r}{ }  & \multicolumn{1}{c}{mas}                         & \multicolumn{1}{c}{mas yr$^{-1}$}        & \multicolumn{1}{c}{mas}                      & \multicolumn{1}{c}{mas yr$^{-1}$} & \multicolumn{1}{c}{km s$^{-1}$} & \\
\multicolumn{1}{r}{(1)} &\multicolumn{1}{c}{(2)}                            & \multicolumn{1}{c}{(3)}                      & \multicolumn{1}{c}{(4)}                        & \multicolumn{1}{c}{(5)}               & \multicolumn{1}{c}{(6)}         & \multicolumn{1}{c}{(7)} \\
\endfirsthead
  \hline
  \endhead
\hline
  \endfoot
  \hline
\multicolumn{7}{p{450pt}}{(1) maser feature ID, (2) initial R.A. offset ($\Delta \alpha_0 \cos{\delta}$) from $\alpha_{\rm c}=\timeform{21h39m40s.501608}$, (3) R.A. proper motion ($\mu_\alpha \cos{\delta}$), (4) initial Decl offset ($\Delta \delta_0$) from $\delta_{\rm c}=\timeform{+51D20'32''.60108}$, (5) R.A. proper motion ($\mu_\delta$), (6) LSR velocity ($V_{\rm LSR}$), and (7) Detection epochs, where the symbol $*$ indicates epochs the maser feature was not detected. Since maser features ID 19--39 were detected only once, the proper motion in colums (3) and (4) are not calculated.} \\
\endlastfoot
  \hline
 1 & $  74.228 \pm  0.213$ & $-2.226 \pm 0.141$ & $ 206.406 \pm  0.212$ & $-2.201 \pm 0.141$ & $-32.3$ &123456***\\ 
 2 & $  71.831 \pm  0.155$ & $-2.866 \pm 0.083$ & $ 194.856 \pm  0.156$ & $-3.317 \pm 0.083$ & $-35.7$ &123456789\\ 
 3 & $  22.334 \pm  0.155$ & $-2.987 \pm 0.083$ & $-131.761 \pm  0.156$ & $-3.211 \pm 0.083$ & $-42.0$ &123456789\\
 4 & $  40.727 \pm  0.155$ & $-3.260 \pm 0.083$ & $-127.782 \pm  0.156$ & $-3.703 \pm 0.083$ & $-42.8$ &123456789\\
 5 & $ -57.609 \pm  0.470$ & $-4.389 \pm 0.411$ & $   8.706 \pm  0.457$ & $-5.244 \pm 0.398$ & $-43.7$ &*234567**\\
 6 & $ -10.550 \pm  0.470$ & $-2.726 \pm 0.411$ & $  47.374 \pm  0.457$ & $-1.793 \pm 0.398$ & $-47.9$ &1234*****\\
 7 & $ -10.283 \pm  0.273$ & $-2.849 \pm 0.166$ & $  44.553 \pm  0.272$ & $-2.189 \pm 0.166$ & $-49.2$ &1234*****\\
 8 & $ -20.904 \pm  0.070$ & $-2.683 \pm 0.973$ & $-144.552 \pm  0.058$ & $-3.102 \pm 0.495$ & $-45.4$ &12*******\\
 9 & $ -18.711 \pm  0.126$ & $-0.122 \pm 0.891$ & $-144.697 \pm  0.050$ & $-3.045 \pm 0.372$ & $-46.2$ &12*******\\
10 & $  71.756 \pm  0.053$ & $-4.402 \pm 1.207$ & $ 200.915 \pm  0.053$ & $-3.179 \pm 0.679$ & $-32.3$ &*23******\\
11 & $ -13.830 \pm  0.136$ & $-1.273 \pm 0.691$ & $  37.586 \pm  0.148$ & $-2.598 \pm 0.743$ & $-46.6$ &**34*****\\
12 & $ -97.535 \pm  0.051$ & $-1.040 \pm 0.373$ & $-128.411 \pm  0.065$ & $-6.251 \pm 0.454$ & $-39.1$ &****56***\\
13 & $ -48.094 \pm  0.050$ & $-2.858 \pm 0.298$ & $ -17.912 \pm  0.053$ & $-8.050 \pm 0.293$ & $-32.7$ &*****67**\\
14 & $ -96.495 \pm  0.034$ & $-6.743 \pm 2.735$ & $-132.491 \pm  0.037$ & $ 6.003 \pm 3.173$ & $-40.3$ &*****67**\\
15 & $ -16.759 \pm  0.539$ & $-3.202 \pm 1.276$ & $  26.884 \pm  0.534$ & $-3.398 \pm 1.262$ & $-44.9$ &******78*\\
16 & $ -39.055 \pm  0.152$ & $-3.114 \pm 0.709$ & $  24.038 \pm  0.064$ & $-7.305 \pm 0.745$ & $-50.0$ &*****67**\\
17 & $-108.001 \pm  0.207$ & $-0.116 \pm 1.042$ & $-118.739 \pm  0.270$ & $-5.270 \pm 1.407$ & $-39.1$ &*******89\\
18 & $ -45.574 \pm  0.185$ & $-1.696 \pm 0.783$ & $  24.077 \pm  0.169$ & $-4.851 \pm 0.711$ & $-49.2$ &*******89\\
19 & $ -55.938 \pm  0.068$ &                    & $  27.719 \pm  0.055$ &                    & $-41.2$ &1********\\
20 & $ -13.798 \pm  0.076$ &                    & $  40.478 \pm  0.062$ &                    & $-48.7$ &1********\\
21 & $ -13.833 \pm  0.130$ &                    & $  40.094 \pm  0.071$ &                    & $-49.6$ &1********\\
22 & $ -21.679 \pm  0.155$ &                    & $  42.460 \pm  0.070$ &                    & $-44.9$ &*2*******\\
23 & $ -39.733 \pm  0.124$ &                    & $ -28.275 \pm  0.061$ &                    & $-43.7$ &***4*****\\
24 & $ -36.587 \pm  0.033$ &                    & $  27.239 \pm  0.022$ &                    & $-50.0$ &***4*****\\
25 & $ -57.389 \pm  0.030$ &                    & $  19.943 \pm  0.032$ &                    & $-34.0$ &*****6***\\
26 & $  69.171 \pm  0.085$ &                    & $ 193.120 \pm  0.106$ &                    & $-36.9$ &*****6***\\
27 & $  20.905 \pm  0.036$ &                    & $-152.029 \pm  0.038$ &                    & $-41.2$ &*****6***\\
28 & $ -15.774 \pm  0.057$ &                    & $  33.159 \pm  0.126$ &                    & $-46.6$ &*****6***\\
29 & $ -18.188 \pm  0.041$ &                    & $  26.321 \pm  0.089$ &                    & $-47.5$ &********9\\
30 & $ -41.598 \pm  0.068$ &                    & $  77.481 \pm  0.052$ &                    & $-25.6$ &******7**\\
31 & $  69.312 \pm  0.100$ &                    & $ 204.313 \pm  0.046$ &                    & $-29.4$ &******7**\\
32 & $ -16.799 \pm  0.146$ &                    & $  26.571 \pm  0.064$ &                    & $-47.1$ &******7**\\
33 & $ -16.135 \pm  0.096$ &                    & $  39.354 \pm  0.187$ &                    & $-48.7$ &******7**\\
34 & $ -16.316 \pm  0.302$ &                    & $  27.892 \pm  0.395$ &                    & $-53.0$ &******7**\\
35 & $ -56.108 \pm  0.170$ &                    & $  23.175 \pm  0.138$ &                    & $-37.8$ &*******8*\\
36 & $ -44.129 \pm  0.067$ &                    & $  70.453 \pm  0.088$ &                    & $-46.6$ &*******8*\\
37 & $ -42.198 \pm  0.107$ &                    & $  18.659 \pm  0.202$ &                    & $-50.0$ &*******8*\\
38 & $ -99.698 \pm  0.069$ &                    & $-136.305 \pm  0.033$ &                    & $-41.2$ &********9\\
39 & $ -18.684 \pm  0.110$ &                    & $  32.914 \pm  0.089$ &                    & $-47.1$ &********9\\
\end{longtable}
%\end{sidewaystable}

\begin{table*}
  \caption{Parallax and Proper Motion}\label{parallax_spot}
  \begin{center}
    \begin{tabular}{rrrrrrr}
      \hline
      \hline
ID    &  $V_{\rm LSR}$ & Parallax          & $\Delta \alpha_0\cos{\delta}$  & $\mu_\alpha\cos{\delta}$ & $\Delta \delta_0$          & $\mu_\delta$      \\
      & (km s$^{-1}$)  & (mas)             & (mas)                          & (mas yr$^{-1}$)        &  (mas)                     & (mas yr$^{-1}$) \\
(1)  & (2)            & (3)               & (4)                            & (5)                      &  (6)                       & (7)               \\
      \hline
1a    & $-31.5$        & $0.201 \pm 0.152$ & $74.392 \pm 0.478$             & $-2.377 \pm 0.356$       & $206.438 \pm 0.480$        & $-2.048 \pm 0.356$\\
1b    & $-31.9$        & $0.256 \pm 0.129$ & $74.310 \pm 0.407$             & $-2.308 \pm 0.303$       & $206.295 \pm 0.408$        & $-1.993 \pm 0.303$\\
1c    & $-32.3$        & $0.318 \pm 0.108$ & $74.212 \pm 0.286$             & $-2.237 \pm 0.189$       & $206.069 \pm 0.283$        & $-2.016 \pm 0.190$\\
2a    & $-34.8$        & $0.222 \pm 0.172$ & $71.284 \pm 0.802$             & $-2.668 \pm 0.345$       & $195.872 \pm 0.811$        & $-3.656 \pm 0.354$\\
2b    & $-35.3$        & $0.243 \pm 0.092$ & $71.839 \pm 0.229$             & $-2.880 \pm 0.123$       & $194.772 \pm 0.239$        & $-3.241 \pm 0.126$\\
2c    & $-35.7$        & $0.264 \pm 0.091$ & $71.823 \pm 0.226$             & $-2.842 \pm 0.121$       & $194.807 \pm 0.235$        & $-3.355 \pm 0.124$\\
2d    & $-36.1$        & $0.257 \pm 0.111$ & $71.876 \pm 0.304$             & $-2.844 \pm 0.157$       & $194.799 \pm 0.320$        & $-3.381 \pm 0.161$\\
3a    & $-41.6$        & $0.381 \pm 0.091$ & $22.263 \pm 0.270$             & $-2.903 \pm 0.174$       & $-132.067\pm 0.270$        & $-2.953 \pm 0.174$\\
3b    & $-42.0$        & $0.263 \pm 0.097$ & $22.315 \pm 0.240$             & $-2.980 \pm 0.128$       & $-131.789\pm 0.250$        & $-3.201 \pm 0.132$\\
4a    & $-42.8$        & $0.274 \pm 0.100$ & $40.686 \pm 0.249$             & $-3.242 \pm 0.133$       & $-127.820\pm 0.259$        & $-3.692 \pm 0.137$\\
5a    & $-48.7$        & $0.196 \pm 0.119$ & $-10.302\pm 0.414$             & $-2.838 \pm 0.253$       & $ 44.636 \pm 0.409$        & $-2.325 \pm 0.246$\\
5b    & $-49.2$        & $0.263 \pm 0.111$ & $-10.358\pm 0.387$             & $-2.793 \pm 0.237$       & $ 44.948 \pm 0.383$        & $-2.495 \pm 0.231$\\
      \hline
Combined &             & $0.262 \pm 0.031$ &                                & $-2.741 \pm 0.082$       &                            & $-2.865 \pm 0.180$\\
      \hline
      \hline
    \end{tabular}
  \end{center}
{(1) maser spot ID, (2) LSR velocity, (3) obtained parallax, (4) initial R.A. offset ($\Delta \alpha_0 \cos{\delta}$) from $\alpha_{\rm c}=\timeform{21h39m40s.501608}$, (5) R.A. proper motion ($\mu_\alpha \cos{\delta}$), (6) initial Decl offset ($\Delta \delta_0$) from $\delta_{\rm c}=\timeform{+51D20'32''.60108}$, and (7) R.A. proper motion ($\mu_\delta$), (6) radial velocity ($V_{\rm r}$). The last row shows the result of combined-fitting, which is the annual parallax fitting made for all maser spots simultaneously assuming the common distance.}  
\end{table*}

\begin{table*}
  \caption{Internal Motions}\label{internal_motion}
  \begin{center}
    \begin{tabular}{r|rr|rr|r}
      \hline
      \hline
ID    & \multicolumn{2}{c}{R.A.}                             & \multicolumn{2}{c}{Decl}                               & $V_{\rm LSR}$  \\
      & ${\mu_\alpha\cos{\delta_{\rm c}}}-\overline{{\mu_\alpha\cos{\delta_{\rm c}}}}$& $V_{\alpha}-\overline{V_{\alpha}}$ & ${\mu_\delta}-\overline{{\mu_\delta}}$ & $V_{\delta}-\overline{V_{\delta}}$ &   \\
      & (mas yr$^{-1}$)                     & (km s$^{-1}$)& (mas yr$^{-1}$)                      & (km s$^{-1}$)& (km s$^{-1}$) \\
      \hline
    1 & $ 0.52 \pm 0.27$ & $  9.3 \pm  5.0$ & $ 0.66 \pm 0.23$ &  $ 12.0 \pm  4.1$ & $-32.3 \pm 0.4$\\
    2 & $-0.13 \pm 0.30$ & $ -2.3 \pm  5.4$ & $-0.45 \pm 0.20$ &  $ -8.1 \pm  3.7$ & $-35.7 \pm 0.4$\\
    3 & $-0.25 \pm 0.30$ & $ -4.5 \pm  5.4$ & $-0.34 \pm 0.21$ &  $ -6.2 \pm  3.7$ & $-42.0 \pm 0.4$\\
    4 & $-0.52 \pm 0.32$ & $ -9.4 \pm  5.7$ & $-0.84 \pm 0.20$ &  $-15.1 \pm  3.6$ & $-42.8 \pm 0.4$\\
    5 & $-1.64 \pm 0.48$ & $-29.5 \pm  8.6$ & $-2.38 \pm 0.21$ &  $-42.9 \pm  3.8$ & $-43.7 \pm 0.4$\\
    6 & $ 0.03 \pm 0.56$ & $  0.5 \pm 10.2$ & $ 1.07 \pm 1.54$ &  $ 19.4 \pm 27.9$ & $-47.9 \pm 0.4$\\
    7 & $-0.11 \pm 0.41$ & $ -2.0 \pm  7.3$ & $ 0.68 \pm 0.30$ &  $ 12.2 \pm  5.4$ & $-49.2 \pm 0.4$\\
    8 & $ 0.07 \pm 0.98$ & $  1.2 \pm 17.6$ & $-0.25 \pm 0.53$ &  $ -4.6 \pm  9.5$ & $-45.4 \pm 0.4$\\
    9 & $ 2.63 \pm 0.89$ & $ 47.5 \pm 16.2$ & $-0.20 \pm 0.41$ &  $ -3.5 \pm  7.5$ & $-46.2 \pm 0.4$\\
   10 & $-1.64 \pm 1.21$ & $-29.6 \pm 21.9$ & $-0.31 \pm 0.70$ &  $ -5.6 \pm 12.7$ & $-32.3 \pm 0.4$\\
   11 & $ 1.47 \pm 0.70$ & $ 26.6 \pm 12.6$ & $ 0.29 \pm 0.76$ &  $  5.2 \pm 13.8$ & $-46.6 \pm 0.4$\\
   12 & $ 1.70 \pm 0.38$ & $ 30.7 \pm  6.9$ & $-3.41 \pm 0.49$ &  $-61.5 \pm  8.8$ & $-39.1 \pm 0.4$\\
   13 & $-0.10 \pm 0.31$ & $ -1.8 \pm  5.6$ & $-5.19 \pm 0.34$ &  $-93.7 \pm  6.2$ & $-32.7 \pm 0.4$\\
   14 & $-3.98 \pm 2.74$ & $-72.0 \pm 49.4$ & $ 8.86 \pm 3.18$ &  $160.1 \pm 57.4$ & $-40.3 \pm 0.4$\\
   15 & $-0.46 \pm 1.28$ & $ -8.3 \pm 23.1$ & $-0.52 \pm 1.27$ &  $ -9.4 \pm 23.0$ & $-44.9 \pm 0.4$\\
   16 & $-0.35 \pm 0.71$ & $ -6.4 \pm 12.9$ & $-4.45 \pm 0.77$ &  $-80.3 \pm 13.8$ & $-50.0 \pm 0.4$\\
   17 & $ 2.61 \pm 1.05$ & $ 47.1 \pm 18.9$ & $-2.42 \pm 1.42$ &  $-43.7 \pm 25.6$ & $-39.1 \pm 0.4$\\
   18 & $ 1.03 \pm 0.79$ & $ 18.6 \pm 14.2$ & $-2.00 \pm 0.73$ &  $-36.1 \pm 13.2$ & $-49.2 \pm 0.4$\\

      \hline
     \hline
    \end{tabular}
  \end{center}
\end{table*}

\begin{table*}
  \caption{3-D velocity in the Cylindrical Coordinate Adopting Some Possible Galactic Constants}\label{3D_velocity}
  \begin{center}
    \begin{tabular}{lllllll}
      \hline
      \hline
$R_0$           & $V_0$       & $R$            & $V_{R_{\rm in}}$ & $V_\theta$  & $V_z$          & Reference\\
kpc             & km s$^{-1}$ & kpc            & km s$^{-1}$      & km s$^{-1}$ & km s$^{-1}$    & \\
      \hline
$8.05\pm 0.45$  & $238\pm 14$ & $9.22\pm 0.43$ & $-3.8\pm 16.9$   & $218\pm 19$ & $1.91\pm 1.70$ & \citet{hon12}\\
8.5             & 220         & $9.63\pm 0.13$ & $ 7.5\pm 13.5$   & $202\pm  7$ & $1.91\pm 1.70$ & \citet{ker86}\\
8.0             & 217         & $9.18\pm 0.17$ & $ 4.4\pm 13.7$   & $199\pm  7$ & $1.91\pm 1.70$ & \citet{deh98a}\\
$8.34\pm 0.16$  & $240\pm  8$ & $9.49\pm 0.20$ & $-1.9\pm 14.9$   & $220\pm 11$ & $1.91\pm 1.70$ & \citet{rei14}\\
      \hline
     \hline
    \end{tabular}
  \end{center}
\end{table*}

\begin{table*}
  \caption{Summary of Adopted and Derived Physical Parameters}\label{summary}
  \begin{center}
    \begin{tabular}{ll}
      \hline
      \hline
      Phase Tracking Center     & $(\alpha, \delta)=(\timeform{21h39m40s.55}, \timeform{+51D20'34''.0})$ (J2000.0) \\
                                & $(l, b) = (95\fdg29670, -0\fdg93660)$           \\
      Center of Maser Map       & $(\alpha_{\rm c}, \delta_{\rm c})=(\timeform{21h39m40s.501608}, \timeform{+51D20'32''.60108})$ (J2000.0)\\
                                & $(l_{\rm c}, b_{\rm c}) = (95\fdg29635672, -0\fdg93695671)$           \\
      LSR Velocity              & $V_{\rm LSR}=-42.27 \pm 0.22$ km s$^{-1}$\\
      Heliocentric Velocity     & $V_{\rm Hel}=-56.43 \pm 0.22$ km s$^{-1}$\\
      Annual Parallax           & $\varpi=0.262\pm 0.031$ mas \\
      Trigonometric Distance from the Sun   & $D=3.82^{+0.51}_{-0.41}$ kpc \\
      The Galactic constants    & $R_0=8.05\pm 0.45$ kpc,  $V_0=238\pm 14$ km s$^{-1}$$^\dagger$ \\
      3-D Solar Motion in LSR   & $(U_\odot, V_\odot, W_\odot)=(10.0, 12.0, 7.2)$ km s$^{-1}$$^\dagger$\\
      Galactocentric Distance   & $R=9.22\pm0.43$ kpc \\
      Proper Motion             & $(\mu_\alpha\cos{\delta},\mu_\delta)=(-2.74 \pm 0.08, -2.87 \pm 0.18)$ mas yr$^{-1}$\\
                                & $(\mu_l \cos{b}, \mu_b)=(-3.95\pm 0.18, -0.34\pm 0.08)$ mas yr$^{-1}$\\
                                & $(V_l, V_b)=(-71.5\pm 10.1, -6.21\pm 1.70)$ km s$^{-1}$\\
      3-D Velocity Vector       & $(U, V, W)=(86.4\pm 10.1, -37.7\pm 1.0, 1.92\pm 1.70)$ km s$^{-1}$\\
                                & $(V_{R_{\rm in}}, V_\theta, V_z)=(-3.8\pm 16.9, 218\pm 19, 1.92\pm 1.70)$ km s$^{-1}$\\
      Bolometric Luminosity of the YSO & $(2.15\pm 0.54)\times 10^3 L_\odot$\\				
      Spectral Type of the YSO  & B2--B3\\
      \hline
     \hline
    $^\dagger$ \citet{hon12} & \\
    \end{tabular}\\
  \end{center}
\end{table*}
\end{document}